\documentclass[graybox]{svmult}

\usepackage{type1cm}        % activate if the above 3 fonts are
\usepackage{makeidx}         % allows index generation
\usepackage{graphicx}        % standard LaTeX graphics tool
\usepackage{multicol}        % used for the two-column index
\usepackage[bottom]{footmisc}% places footnotes at page bottom

\usepackage{newtxtext}       %
\usepackage{newtxmath}       % selects Times Roman as basic font
\usepackage{siunitx}
\newcommand{\EXCLUDE}[1]{}

\makeindex             % used for the subject index
\begin{document}

\title*{Set~Voronoi Tessellation for Particulate Systems in Two Dimensions}
\author{Simeon V\"olkel \and Kai Huang}
\institute{Simeon V\"olkel \at University of Bayreuth, Experimental Physics V, Universit\"atsstra\ss e 30, 95447 Bayreuth, Germany, \email{simeon.voelkel@uni-bayreuth.de}
\and Kai Huang \at Duke Kunshan University, Division of Natural and Applied Sciences, No. 8 Duke Avenue, Kunshan, Jiangsu, China 215316, \email{kai.huang186@dukekunshan.edu.cn} \&
\at University of Bayreuth, Experimental Physics V, Universit\"atsstra\ss e 30, 95447 Bayreuth, Germany}
\maketitle

\abstract{%
Given a countable set of points in a continuous space,
Voronoi tessellation is an intuitive way of partitioning the space
according to the distance to the individual points.
As a powerful approach to obtain structural information,
it has a long history and widespread applications in diverse disciplines,
from astronomy to urban planning.
For particulate systems in real life,
such as a pile of sand or a crowd of pedestrians,
the realization of Voronoi tessellation needs to be modified to accommodate the fact
that the particles cannot be simply treated as points.
Here, we elucidate the use of Set~Voronoi tessellation
(i.\,e., considering for a non-spherical particle a \emph{set of points} on its surface)
to extract meaningful local information in a quasi-two-dimensional system of granular rods.
In addition, we illustrate
how it can be applied to arbitrarily shaped particles
such as an assembly of honey bees or pedestrians
for obtaining structural information.
Details on the implementation of this algorithm
with the strategy of balancing computational cost and accuracy
are discussed.
Furthermore,
we provide our python code as open source
in order to facilitate Set~Voronoi calculations
in two dimensions for arbitrarily shaped objects.
}

\section{Introduction}
\label{sec:introduction}
\begin{figure}[tb]
\sidecaption[t]
\includegraphics[trim={0 2.0cm 0 0},clip,width=7.5cm]{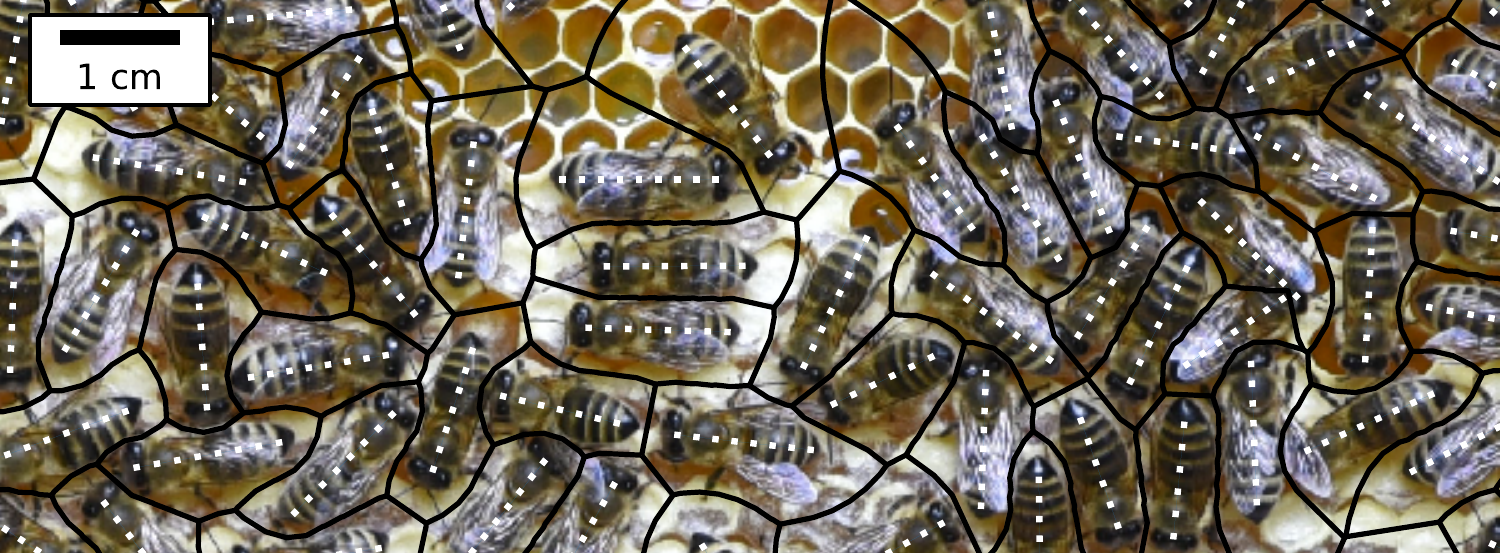}
\caption{%
Carniolan honey bees on a partially sealed honeycomb.
%The photo was taken on 2019-09-14 at 09:08 local time in Bayreuth, Germany.
The Set~Voronoi tessellation (black) is based on multiple points (white) per bee.
}
\label{fig:bees}
\end{figure}
Particulate systems are ubiquitous in nature, industry and our daily lives,
ranging from active ones
like pedestrians or animals, as in Fig.~\ref{fig:bees},
to passive ones such as athermal granules.
\EXCLUDE{
From cereals at home to
the proverbial sand of the sea in nature,
the the latter class is quasi omnipresent.
}
They all
share the characteristics
that the macroscopic behavior depends on
the arrangement of individual particles relative to each other.
Prominent collective effects regarding the dynamics of granular materials,
such as the solid-liquid-like transition \cite{Huang2010, Huang2009, Huang2009a, Andreotti2013, May2013, Zhao2014, Ramming2017, Baur2017}
and pattern formation \cite{Ristow2000, Fortini2015, Zippelius2017},
are typically triggered by the mobility of individual particles.
Moreover,
there also exists evidence showing that the strain field associated with local rearrangement of particles
can be used to obtain the local stress field,
which in turn can provide indispensable hints
on the establishment of local force networks during jamming transition \cite{Liu1998,Bi2011,Zhao2019}.
Investigations on granular systems thus often rely on accurate measurements of properties
such as local volume fraction, neighborhood, etc.
For defining these quantities,
it is expedient to attribute to every particle a portion of the available space,
that ``belongs to'' or ``is occupied'' by a single particle.

\section{Limitations of the classical Voronoi tessellation}

The honeycomb in Fig.~\ref{fig:bees} can be seen as an example of a naturally formed tessellation,
optimized to have a fair distribution of space for individual larvae of bees to grow inside.
Motivated by such self-organized processes in nature,
the concept of spatial tessellation was established long time ago \cite{Descartes1644, Dirichlet1850, Voronoi1907, Okabe2000}
in order to analyze structures of various systems in diverse disciplines,
from the structure of the universe in astronomy \cite{Descartes1644}
through positioning public schools or post offices in urban planning \cite{McAllister1996}
to characterizing topological aspects of molecular structures \cite{Mackay1995,Aurenhammer1991,Bernauer2008}.

A generic, parameterless approach
to attribute space based on a single point per object
(typically its center)
is the Voronoi tessellation (VT),
also known as Dirichlet tessellation, Thiessen polygons or Wigner-Seitz cells \cite{Dirichlet1850, Voronoi1907, Thiessen1911, Wigner1933}.
\begin{figure}[b]
%\centering
\sidecaption[t]
\scalebox{0.68}{
\newlength{\vshift}
\setlength{\vshift}{0.27\textwidth}
%\addtolength{\vshift}{-\baselineskip}
\raisebox{\vshift}{\large a)\hspace*{-2em}}\raisebox{\baselineskip}{\includegraphics[height=0.27\textwidth]{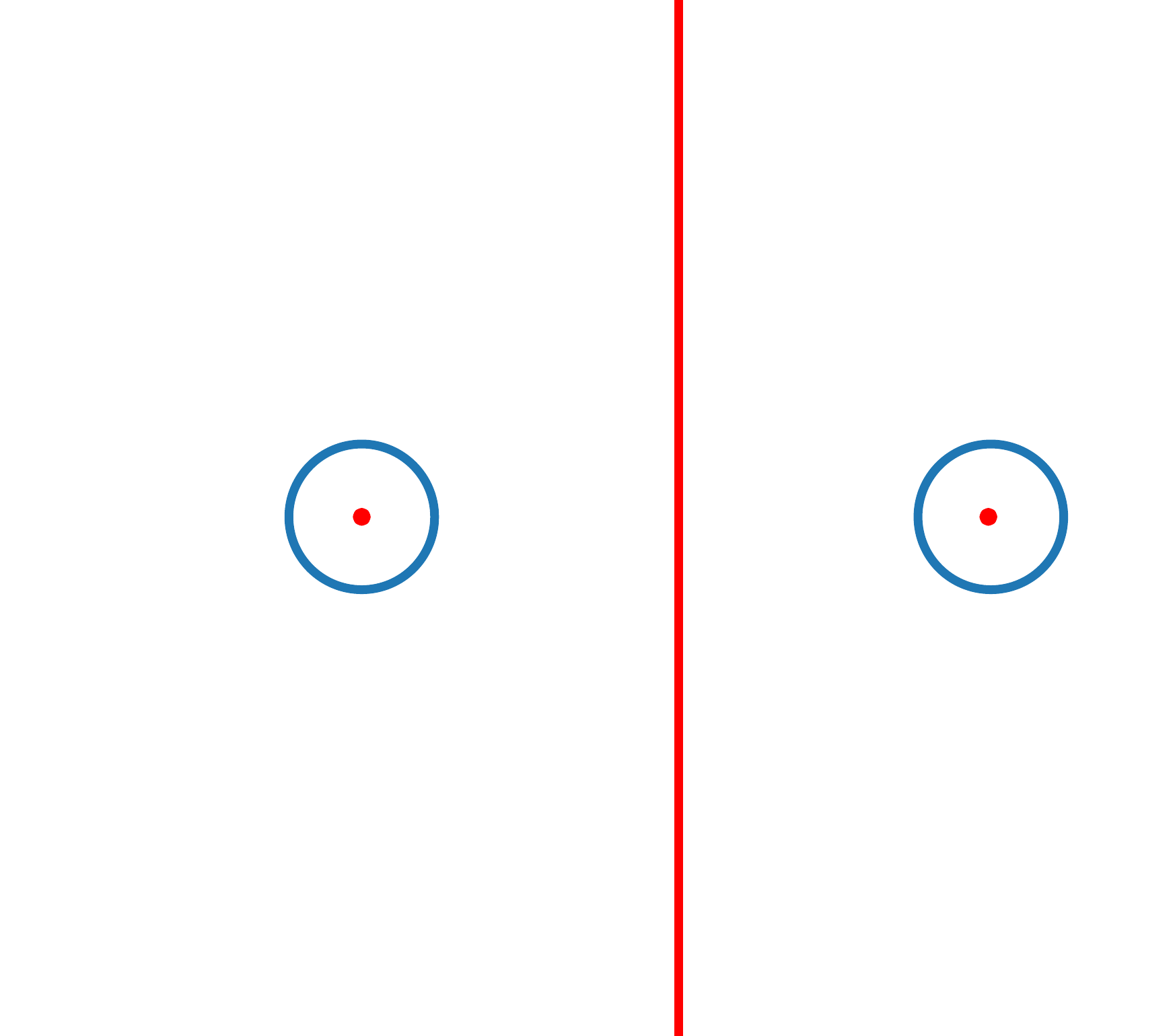}}
\hfill
\raisebox{\vshift}{\large b)\hspace*{-2em}}\raisebox{\baselineskip}{\includegraphics[height=0.27\textwidth]{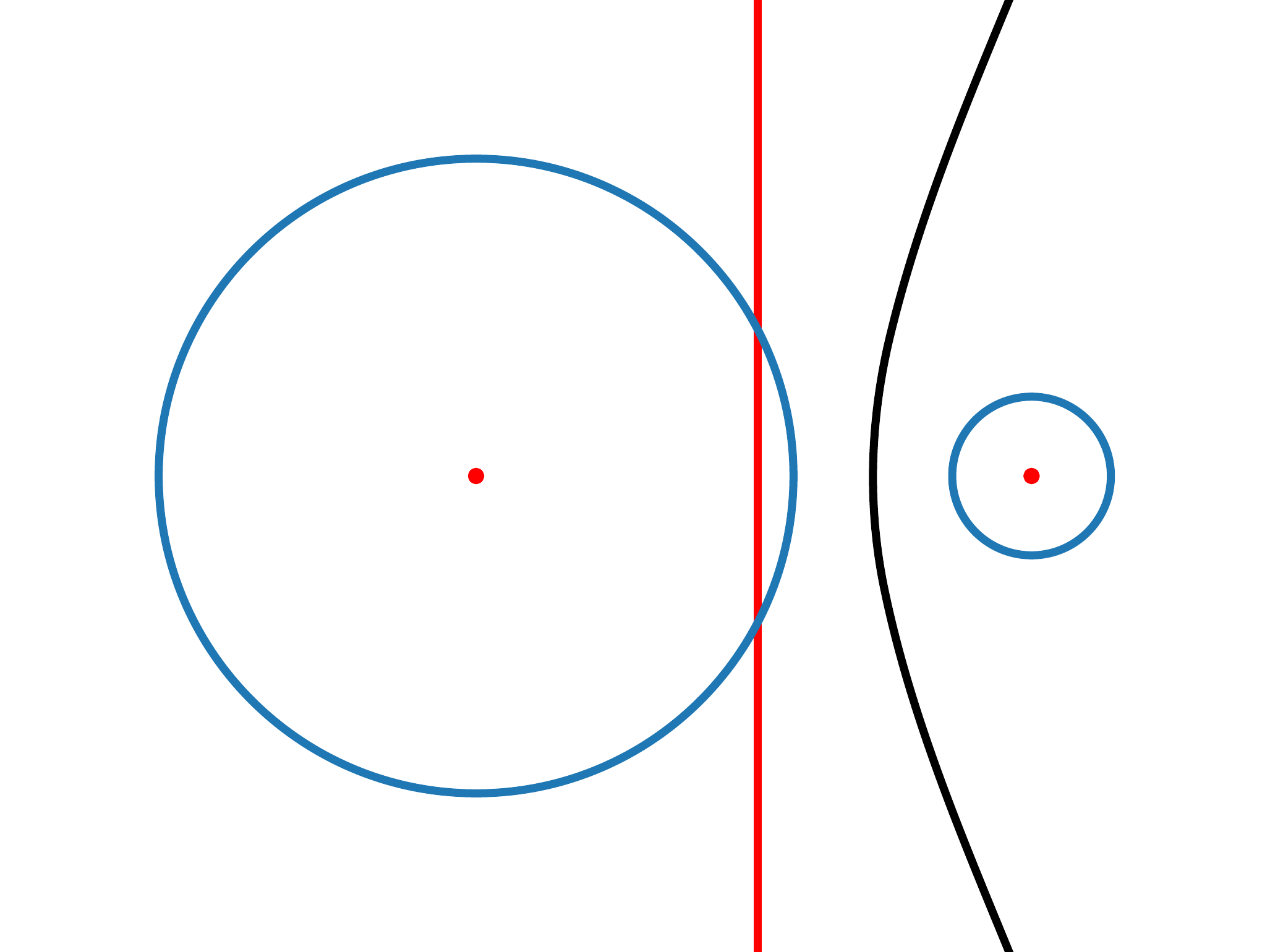}}
\hfill
\raisebox{\vshift}{\large c)\hspace*{-2em}}\raisebox{\baselineskip}{\includegraphics[height=0.27\textwidth]{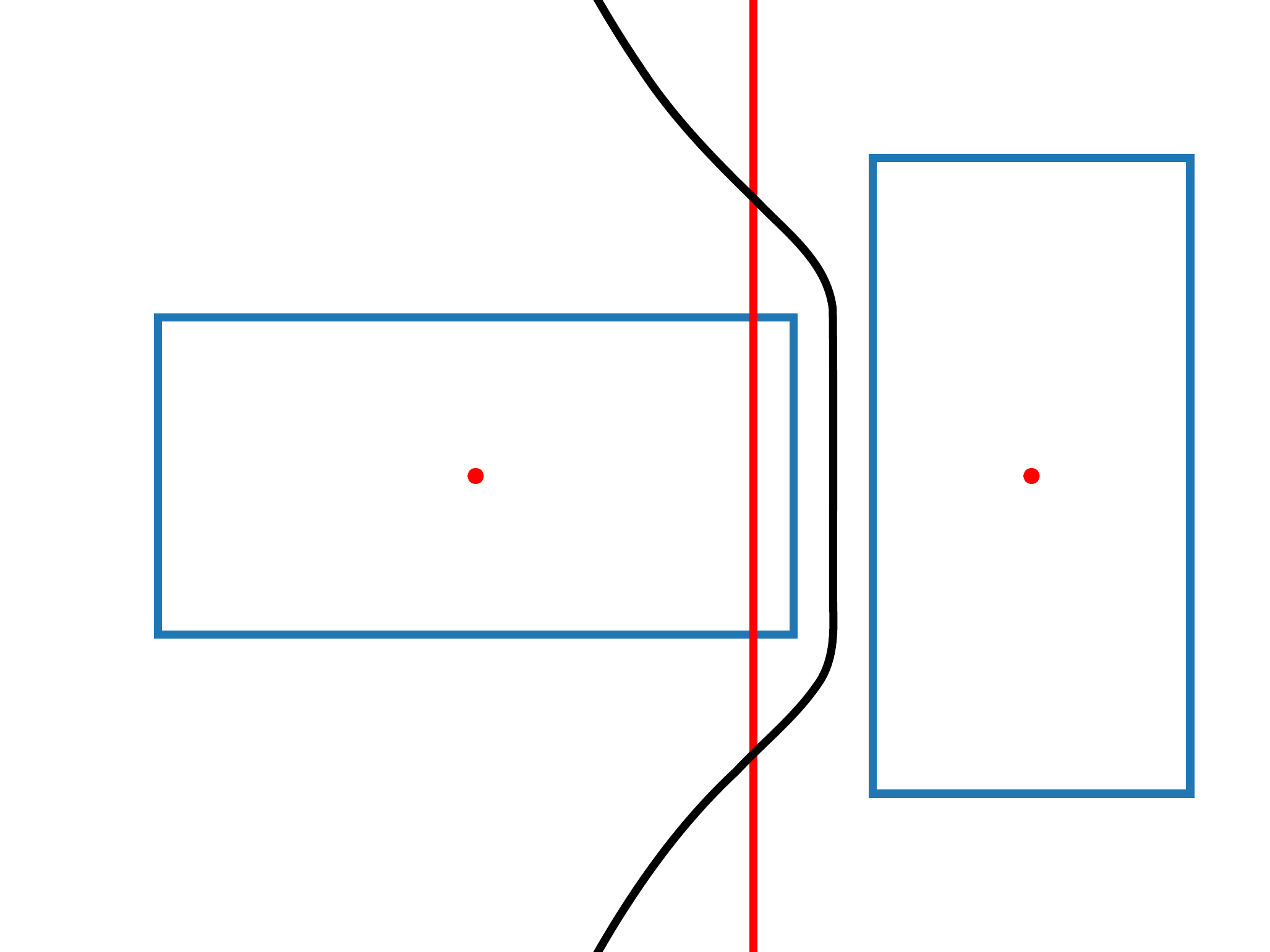}}
}
\caption{
Comparison of classical VT based on the center (red) of particles and
Set~VT (black) considering the border of the two particles (blue).
}
\label{fig:tessellations}
\end{figure}

For mono-disperse spherical particles
(as well as point-like particles),
classical VT based on the particle centers
is applicable and delivers intuitive results
(see Fig.~\ref{fig:tessellations}a).
This, however, is not the case for polydisperse packings (see Fig.~\ref{fig:tessellations}b)
or non-spherical particles (see Fig.~\ref{fig:tessellations}c).
Here, part of the particle on the left lies outside of the Voronoi cell,
contributing to an obvious source of error.
To overcome this,
various weighted Voronoi diagrams have been proposed (see, e.\,g., \cite{Okabe2000} for an overview)
and implemented in physics and material science \cite{Park2007, Park2012}.

Problems due to applying classical VT
to a glass composed of \emph{differently sized} (but spherical) atoms
have already been discussed \cite{Park2012}.
Here, we highlight the problems
arising from applying classical VT to equally sized, but \emph{elongated} particles.

\section{Set~Voronoi algorithm and implementations}

A generic way to attribute space to both
differently sized particles and non-spherical particles
is to tessellate space according to the distance to the closest \emph{surface}
(as opposed to center).
This can be seen as the limiting case of a classical VT,
when considering for each particle the \emph{set of points} marking its surface.
Therefore it is also referred to as \emph{Set~Voronoi} diagram \cite{Schaller2013}.
The resulting, intuitive cell-borders of the examples in Fig.~\ref{fig:tessellations}
are depicted in black.
Note that the Set~Voronoi border for nonoverlapping monodisperse spheres
(or circular discs in two dimensions)
like in Fig.~\ref{fig:tessellations}a
coincides with the classical Voronoi cell border.
Voronoi diagrams with lines and arcs as generators have been studied systematically since the late 1970s
\cite{Okabe2000}.
Nevertheless, their importance in the realm of granular physics was rarely realized until the past decade \cite{Baule2013,Schaller2013,Baule2014,Weis2017}.

Recently,
Weis \& Sch\"{o}nh\"{o}fer \cite{Pomelo} provided a program
based on the Voro++ library \cite{Rycroft2009}
to calculate the Set~VT, % in three dimensions,
targeting reconstructed CT-scans of three-dimensional (3D) packings of non-spherical granular particles \cite{Weis2017}.
However,
an urgent demand for a solution in two dimensions remained,
as it is computationally not economic to apply the 3D algorithm directly
to two-dimensional (2D) systems \cite{Weis_PrivateDiscussion}:
Adding dimensions to the problem inevitably increases the computational cost for obtaining the tessellation,
letting aside the efforts now required to treat the borders in the added dimension(s).
Considering as well the fact that many particulate problems of practical interest
(e.\,g., dynamics of pedestrians, flocks or monolayers of granulates)
can be treated (quasi-) two-dimensionally
and very often data is acquired using 2D imaging techniques,
we implemented the Set~VT strategy based on discrete points
put forward by Schaller et al. \cite{Schaller2013}
in 2D.
Our implementation
uses the python scripting language
together with scipy/numpy libraries \cite{scipy},
relying on VT routines provided by the Qhull library \cite{Barber1996}.
It is available as free open source software \cite{git_link}.

\section{Granular rod monolayer as a test case}

\begin{figure}[b]
\centering
\scalebox{0.75}{
\begin{minipage}[b]{0.8\textwidth}
%\newlength{\vshift}
\setlength{\vshift}{0.34\textwidth}
\raisebox{\vshift}{\large a)\hspace*{-1em}}%
\includegraphics[width=\textwidth]{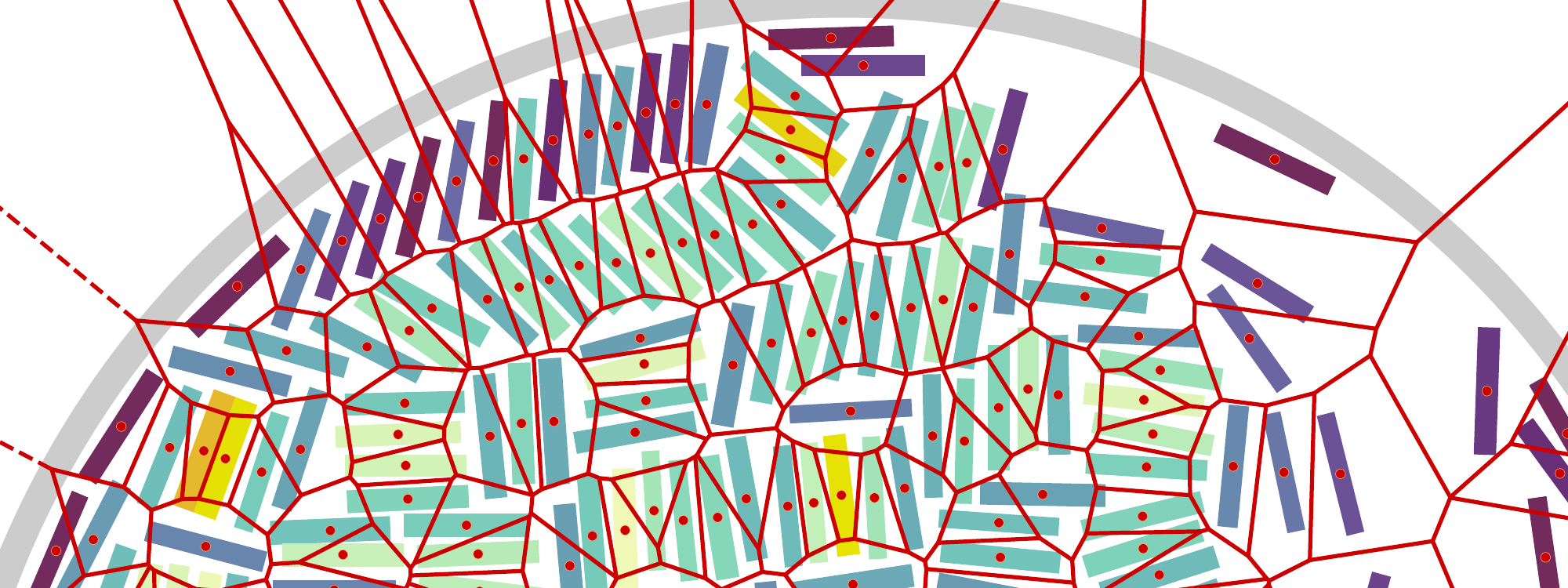}

\addtolength{\vshift}{-0.5\baselineskip}
\raisebox{\vshift}{\large b)\hspace*{-1em}}%
\includegraphics[width=\textwidth]{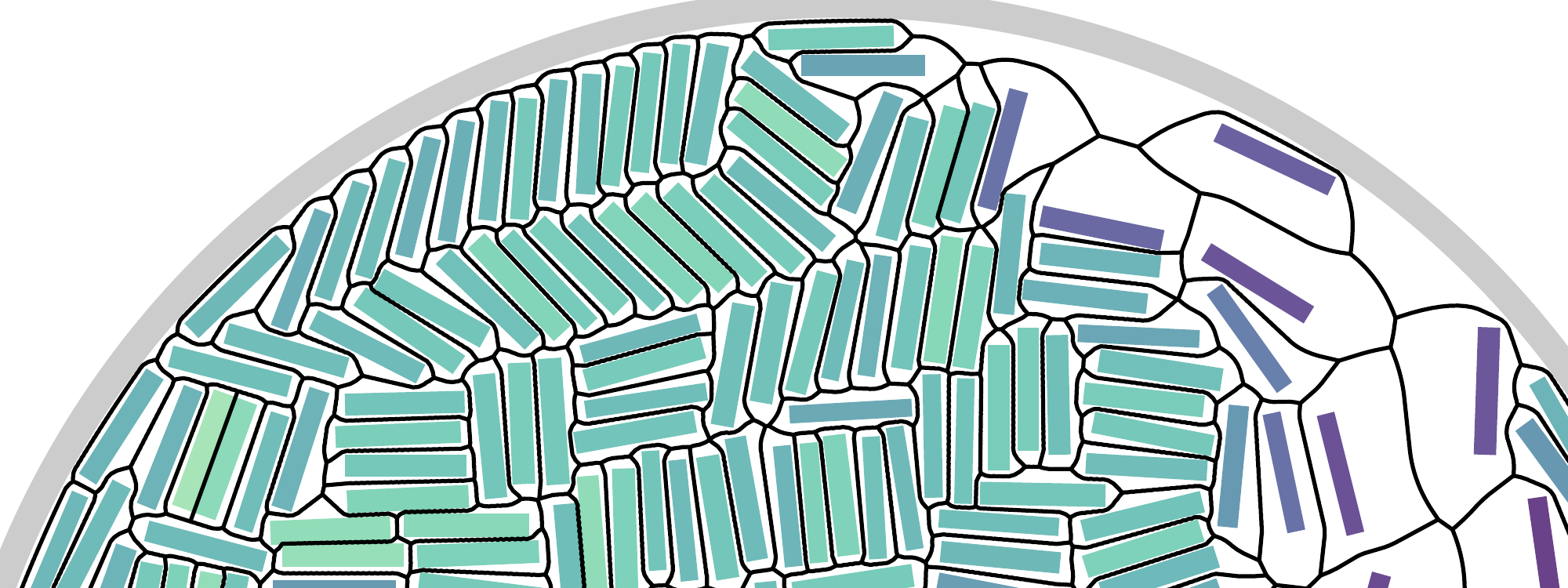}
\end{minipage}
\hfill
\includegraphics[width=0.15\textwidth]{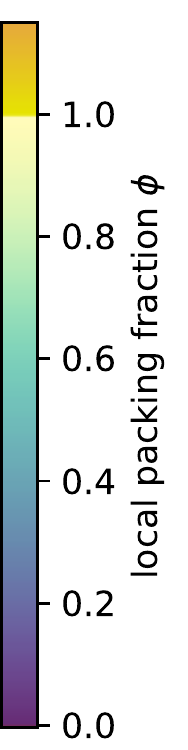}
}
\caption{
VT (a)
and Set~VT (b)
of partly ordered granular rods in quasi-two-dimensions.
The parameters are:
particle number $N=400$,
driving frequency $f = \SI{50}{\hertz}$
and dimensionless acceleration $\Gamma = 6.26$.
The color code indicates the local packing fraction~$\phi$
according to the different tessellations:
classical VT considering the particle centers (red, upper panel)
vs.\ Set~VT (black, lower panel)
including the container rim as additional particle.
}
\label{fig:reconstruction}
\end{figure}

As an example,
we demonstrate the advantages of Set~VT using a monolayer of monodisperse granular rods
of length $l=\SI{15}{\milli\meter}$ and diameter $d=\SI{3}{\milli\meter}$.
They
are confined in a horizontal cylindrical container of diameter $D = \SI{19}{\centi\meter}$,
which is subjected to sinusoidal vibrations against gravity
with oscillation frequency $f$ and dimensionless acceleration $\Gamma$
as two control parameters to keep the rods mobilized.
More details on the experimental set-up and image analysis procedure can be found in \cite{Mueller2015}.
Depending on the packing density,
the rods may organize themselves into an uniaxial nematic state with two-fold rotational symmetry
or tetratic state with 4-fold symmetry.
For analyzing the disorder-order transition quantitatively,
an accurate determination of the local packing density is desired.
For amorphous media,
such as a random packing of granular particles,
(Set) VT provides a direct route
to the local area or volume fractions of individual grains \cite{Song2008, Torquato2010, Schaller2013}.

\nocite{Tange2018}
Fig.~\ref{fig:reconstruction} compares the outcome of classical and Set~VT
based on a snapshot reconstructed from
the positions and orientations obtained experimentally in \cite{Mueller2015}.
This particular snapshot is chosen,
as it consists of both dense and dilute regions.
The classical VT does not include walls and
leads to cells cutting the walls of a container
(red solid lines crossing the gray area in Fig.~\ref{fig:reconstruction}a)
or even extending to infinity (red dashed lines in Fig.~\ref{fig:reconstruction}a).
These problems can be easily and consistently avoided using Set~VT,
as the container can be included as an additional `particle',
naturally limiting the cells of all contained particles (see Fig.~\ref{fig:reconstruction}b).
Additionally Set~VT
delivers a much more reasonable tiling
in the sense that no particle cuts its cell's border,
as the container lid prohibits the `hard' rods from overlapping.
Quantitatively,
the local packing density $\phi = A_\text{p}/A_\text{c}$ derived from the projected area of the particles $A_\text{p} = l \cdot d$
and the area of the corresponding cell $A_\text{c}$ obtained from the space tiling
also demonstrates the difference clearly:
Cells with `impossible' $\phi > 1$ (see color code) disappear
and large fluctuations of $\phi$ for rods aligned with each other in a similar local configuration
diminish as Set~VT is implemented.

\begin{figure}[bt]
\centering
\includegraphics[width=0.7\textwidth]{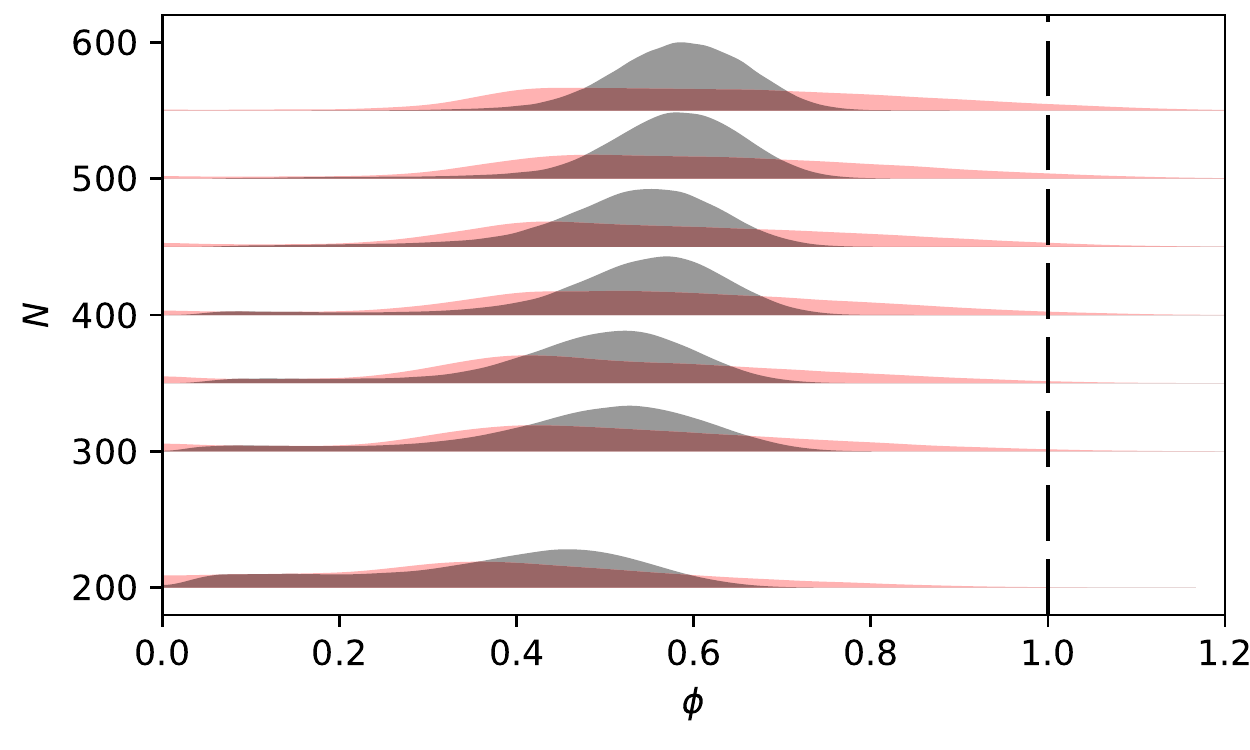}
\caption{%
Probability density $P_\phi$ of the local packing fraction $\phi$ for different numbers of particles $N$ in the container,
obtained from kernel density estimation.
Light red and dark gray curves correspond to classical and Set~VT, respectively.
The peak acceleration is kept in the range $\Gamma \in [1.94; 20.59]$ and $f = \SI{50}{\hertz}$ is fixed.
Other experimental parameters are the same as in Fig.~\ref{fig:reconstruction}.
The reflection method \cite{Silverman1986,Boneva1971} has been used to correct for $\phi$ being non-negative by definition,
and the bandwidth of the Gaussian kernel is chosen
according to Scotts rule \cite{Scott2015}
as $\sigma \cdot n^{-1/5}$ % See equation 6.44 on page 164 in Scott2015
proportional to the standard deviation $\sigma$ of the distribution
and dependent on the number of rods $n$ (here \numrange{1.7e5}{5.3e5} per stacked plot).
Captured frames with detection problems
(particles (partially) outside of the container, overlapping or of zero size)
were skipped to avoid bias.
}
\label{fig:particlenumberdependence}
\end{figure}

Fig.~\ref{fig:particlenumberdependence} compares
the local packing density distribution
obtained from VT (red) and Set~VT (black)
for different global packing densities.
For Set~VT, the probability density $P_{\phi}$
shows a clear trend of an increasing local packing density with the global one.
This feature is much less obvious for the classical VT.
There,
the most striking observation is the tail of the distribution towards large $\phi$ becoming more prominent.
The dashed line at $\phi = 1$ marks the upper bound for hard particles and a reasonable tiling.
All red curves clearly exceed this limit.
This manifests again the importance of observing the applicability of each tessellation technique.

\section{Why Set~Voronoi is essential for elongated particles}

To analyze quantitatively the maximal error
introduced by applying the classical VT to elongated particles,
we consider a perfect, dense packing of identical ``hard'' rectangles.
With no space left between the non-overlapping rectangles,
a constant local packing fraction equal to the global packing fraction of unity is expected.

\begin{figure}[tb]
%\centering
\sidecaption[t]
\hspace*{-1em}\scalebox{0.65}{
\begin{minipage}[b]{0.84\textwidth}
\includegraphics[width=\textwidth]{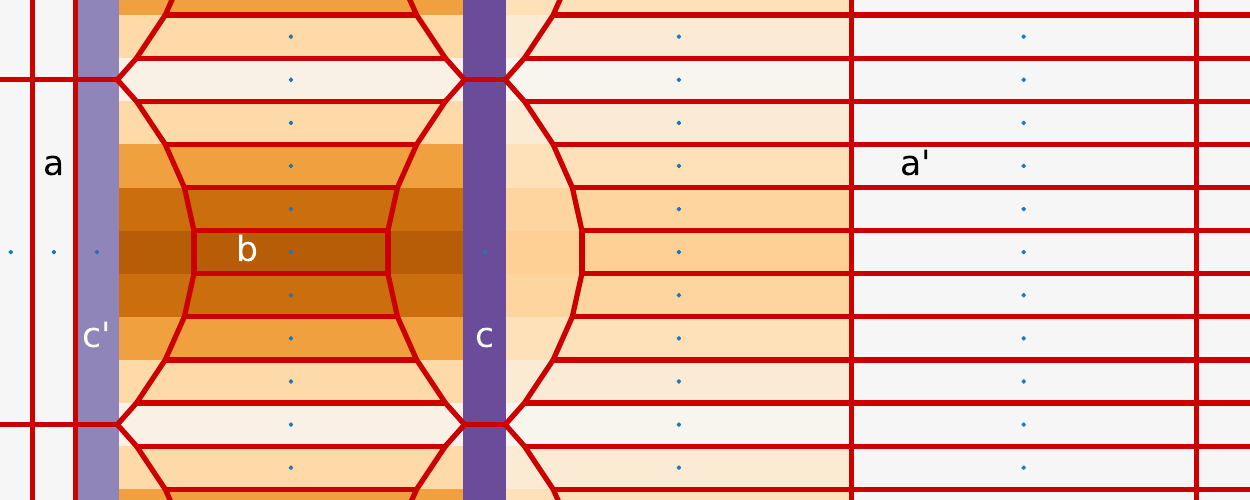}
\end{minipage}
\hfill
\includegraphics[width=0.145\textwidth]{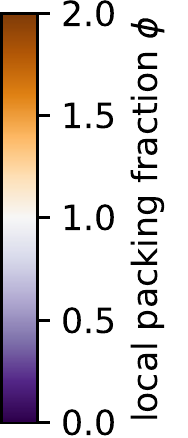}
}
\caption{
Local packing fraction (see color code) according to classical VT (red lines, based on particle centers)
in a dense packing of rectangles with $\varepsilon = 1/8$ exhibiting tetratic ordering.
}
\label{fig:tetratic}
\end{figure}

Fig.~\ref{fig:tetratic} gives an example for such a dense packing of
rectangles with width-to-length ratio $\varepsilon$.
%together with the Voronoi tessellation based on particle centers shown in red.
%The local packing fraction based on classical VT (shown in red)
Classical VT (red lines) gives
the expected value for the local packing fraction,
as indicated by the color code,
if all neighboring particles are aligned in the same direction
(cf.~particles a and a' in Fig.~\ref{fig:tetratic})
but fluctuates significantly
where rectangles of different orientation come together.
Here, local packing fractions
up to $\phi_\text{VT,max} = 2 / (1 + \varepsilon)$ are obtained, e.\,g.,
for particle b in Fig.~\ref{fig:tetratic}.
This overestimation is thus a first order effect in terms of $\varepsilon$.
It can reach a remarkable error of \SI{100}{\percent}
for vanishing width-to-length ratio,
even when letting aside any experimental inaccuracies or elasticity of the particles.
At the same time, the local packing fraction for neighboring particles c and c' in Fig.~\ref{fig:tetratic} is underestimated,
up to an equally dramatic extent as $\varepsilon \rightarrow 0$.
Even for a \emph{dense} packing of hard particles (global packing fraction of unity) and without boundary effects,
classical VT can indicate \emph{vanishing} local packing fraction,
deviating disastrously from the expected value!

On the contrary, Set~VT
gives consistent local packing fractions.
It delivers a value of $\phi_\text{Set\,VT} = 1$ for every rectangle in Fig.~\ref{fig:tetratic}
within the numerical uncertainty arising from the approximation of the surface through a finite number of points.

\section{Performance vs.~quality trade-off via erosion}

For closely spaced particles
special care is required to identify their surface properly
before applying the Set~VT.
For rounded particles with a finite minimal curvature radius $r_\text{c}$,
Schaller et al. \cite{Schaller2013} show that it is highly beneficial to consider the maximally eroded surfaces,
having the constant minimal distance $r_\text{c}$ to the original ones:
Using the eroded surface, dramatically fewer points suffice for obtaining the same Set~VT and accuracy.
In addition,
erosion can resolve slight particle overlap.
This is particularly helpful for densely packed particles close to jamming \cite{Zuriguel2014}
and is expected to make investigations of deformable particles feasible \cite{Schaller2013}.

For the granular rods discussed above,
the rectangles seen by the camera have sharp corners.
In other words, $r_\text{c} = 0$, preventing a `lossless' erosion.
Accurately capturing the sharp corners commands a vanishing erosion depth,
which in turn dictates a very close spacing of the discretization points on the eroded surface.
As describing a surface with a larger number of points inevitably increases the computational costs,
a compromise between accuracy and speed has to be found.
Here, we erode the rods by one pixel
to remove slight overlap due to finite experimental resolution, detection accuracy and finite elasticity of the particles.
Setting the maximum distance between discretization points to the erosion depth delivers satisfactory results
and we recommend this as a rule of thumb.

For many systems, however,
the situation is more pleasant, especially
if the most important property is the elongation,
while the dimensions along the other direction(s) are equal
and the exact shape of each particle only plays a tangential role,
like in the case of the bees shown in Fig.~\ref{fig:bees}.
In such a case,
it is typically acceptable to approximate the particle as spherocylinder.
This approximation paves the way to a very efficient representation of the particle when using Set~VT:
The maximally eroded `surface' of each particle is then just its medial axis,
a one-dimensional line segment.
Furthermore, according to the rule of thumb,
one discretization point per radius of the spherocylinder $r$
suffices for satisfactory results even at high packing densities.
For the bees on the honeycomb,
this translates to a single digit number of discretization points,
each depicted as a white dot in Fig.~\ref{fig:bees}.
Nevertheless,
this is sufficient for resolving the most prominent and important effects due to the elongated shape of the individual bee,
as the black Set~Voronoi cell borders illustrate.

As similarly sized rod-shaped particles represent a diverse class of systems,
from liquid crystal molecules at a microscopic scale to pedestrians viewed from the top (i.\,e., when taking their shoulders into account),
the above analysis demonstrates that it is essential to employ Set~VT for elongated particles.

\section{Conclusion}

Using dynamics of a granular rod monolayer as an example,
we demonstrate that Set~VT provides a more meaningful tiling of space
in comparison to the classical VT
that relies on the center of particles.
From polydisperse systems to irregularly shaped or
even deformable particles,
the Set~VT algorithm is expected to be substantially more consistent
in characterizing the geometric and topological features of particulate systems,
many of which can be approximated as elongated particles.

Note that in addition to obtaining the local packing density,
VT can also be used to extract other order parameters,
such as determining neighbors of individual particles.
The Delaunay triangulation,
which connects particle centers to their neighbors,
follows the natural definition that neighbors share a part of a Voronoi cell border
and is a typical approach after classical VT.
The extension of this definition to Set~VT is straightforward and
can be used in further characterizations,
for instance, using the bond orientational order parameter \cite{May2013}.
How to extract more meaningful information from Set~VT
in addition to the local packing density
and the improvements against the classical VT
will be a focus of future investigations.

\begin{acknowledgement}
We gratefully acknowledge helpful discussions with Simon Weis, Matthias Schr\"{o}ter, Ingo Rehberg, Stefan Hartung and Wolfgang Sch\"{o}pf.
We thank Alexander and Valentin Dichtl
for the possibility of taking the picture of the bees in Fig.~\ref{fig:bees}
during an inspection of the hive.
This work is supported by the German Research Foundation (DFG) under
Grant No. HU1939/4-1.

\end{acknowledgement}

\bibliography{voronoi}

\end{document}